\documentclass{article}

\usepackage{cite}
\usepackage{graphicx}
\usepackage{caption}
\usepackage{amssymb}
\usepackage{array}
\usepackage{dcolumn}
\usepackage{bm}
\usepackage{textgreek}
\usepackage{amsmath}
\usepackage[letterpaper, total={7in, 9in}]{geometry}
\usepackage{authblk}

\title{Optical Isolation with Microring Modulators}

\author[1,*]{Nathan Dostart}
\author[2]{Hayk Gevorgyan}
\author[2]{Deniz Onural}
\author[2]{Milo\v s Popovi\'c}

\affil[1]{Dept. of Electrical, Computer, and Energy Engineering, University of Colorado, Boulder, CO, 80309, USA}
\affil[2]{Dept. of Electrical and Computer Engineering, Boston University, Boston, MA, 02215, USA}

\affil[*]{Corresponding author: nathan.dostart@colorado.edu}

\date{\today}

\begin{document}

\maketitle

\begin{abstract}
	Optical isolators, while commonplace in bulk- and fiber-optic systems, remain a key missing component in integrated photonic systems. Isolation using magneto-optic effects has been difficult to implement due to fabrication restraints, motivating use of other non-reciprocal effects such as temporal modulation. We demonstrate a non-reciprocal modulator comprising a pair of microring modulators and a microring phase shifter in an active silicon photonic process which, in combination with standard frequency filters, facilitates isolation. Isolation up to 13 dB is measured with a 3 dB bandwidth of 2 GHz and insertion loss of 18 dB. As one potential application is cross-talk suppression in bi-directional communication links, we also show transmission of a 4 Gbps data signal through the isolator while retaining a wide-open eye diagram. This compact design, in combination with increased modulation efficiency, could enable modulator-based isolators to become a standard `black-box' component in integrated photonics foundry platform component library. 
\end{abstract}


Optical isolators, devices which allow light to pass in one direction but not the other, are essential components for protecting lasers from reflected light to enable stable operation. The incorporation of gain components, and by extension lasers, into existing silicon-based integrated photonic foundries has made significant progress \cite{komljenovic2016heterogeneous}; soon on-chip lasers will be ubiquitous. Accompanying isolators will be needed to protect these lasers, motivating the further development of on-chip isolators.

An ideal isolator should transmit all the incident light in the desired (forward) direction and no light in the undesired (backward) direction. Isolation requires breaking Lorenz reciprocity \cite{jalas2013and}, which can be achieved with magneto-optic materials, time-varying permittivity, or nonlinear interactions. Silicon lacks magneto-optic properties and magneto-optic materials such as Ye:CIG are exceptionally difficult to integrate with silicon \cite{sounas2017non}, motivating non-magneto-optic approaches to isolation. Time-varying permittivity \cite{sounas2017non} is a particularly appealing alternative as it is linear in the isolated signal and silicon-based optical modulators provide a simple and low risk solution to create this time-varying permittivity. 

Exceptionally few modulator-based isolators have been demonstrated in silicon \cite{lira2012electrically,doerr2014silicon}, while several other interesting approaches have been proposed for silicon photonics but demonstrated in simulation \cite{yu2009complete} or other platforms, particularly off-the-shelf lithium niobate modulators \cite{yang2014experimental,dong2015travelling}. Two main drawbacks can be identified in the silicon photonic modulator-based isolators demonstrated so far: low isolation ($< 5$ dB) \cite{lira2012electrically,doerr2014silicon} and exceptionally long device lengths (1 cm \cite{doerr2014silicon} and 2 cm \cite{lira2012electrically}). These drawbacks both arise from the need to accumulate a certain amount of time delay, at the speed of light, in order to achieve a desired phase shift between the modulated sidebands and carrier. This requires long delay lines (device length) on-chip within which fabrication variations, loss, and other undesired effects degrade performance.

In this paper, we propose and demonstrate a new, resonant isolator design based on silicon microring modulators \cite{xu2005micrometre,sun2019128} which is both compact (300 {\textmu}m $\times$ 50 {\textmu}m here, with the potential to further decrease size) and achieves an isolation of 13.1 dB, nearly 10 dB higher than previous demonstrations in silicon. This design uses a pair of dual ring modulators in a configuration analogous to the isolator in \cite{doerr2014silicon} to generate only a single modulated sideband, rather than the many sidebands generated in the previously studied broadband case. This sideband can therefore be phase-shifted relative to the carrier using only a simple microring phase shifter, wrapping up the delay line into a compact footprint which is less sensitive to fabrication variations.


\begin{figure}[t]
	\centering
	\includegraphics[width=.8\textwidth]{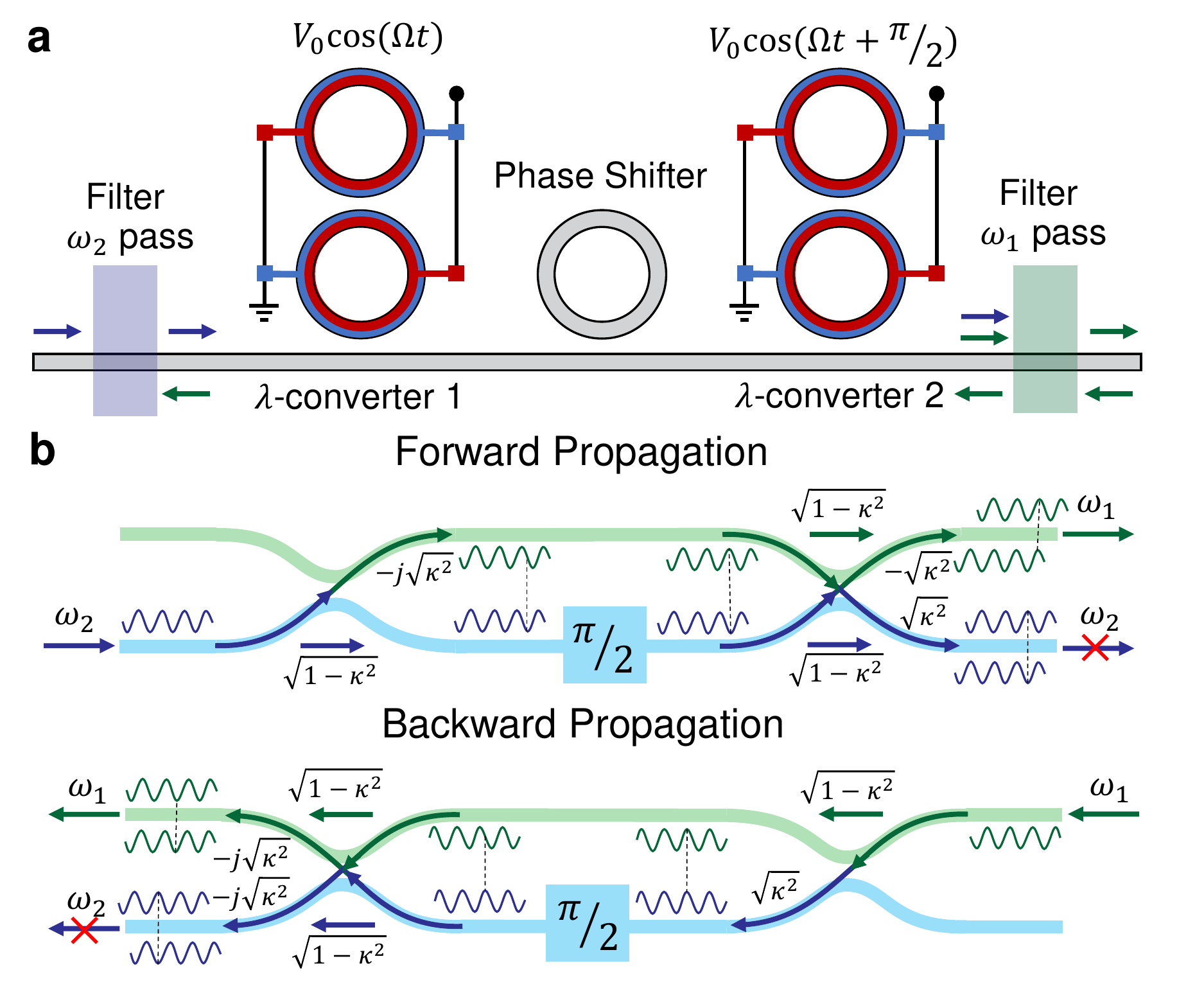}
	\caption{
	\textbf{Isolator concept.}
	\textbf{a} Schematic depiction of isolator, composed of two filters, two wavelength converters, and a microring phase shifter.
	\textbf{b} Mach-Zehnder interferometer analogue where the two arms are the two super mode resonances in the forward (top) and backward (bottom) directions. Each component of the non-reciprocal modulator in \textbf{a} is vertically aligned with the equivalent action in \textbf{b}. In the forward direction light is injected in the bottom arm ($\omega_2$, blue) and exits in the top arm ($\omega_1$, green). In the backwards direction the non-reciprocal phase acquired from wavelength conversion causes the injected light to remain in the top arm, resulting in non-reciprocal wavelength conversion.}
	\label{fig:concept}
\end{figure}

The proposed isolator, schematically shown in Fig.~\ref{fig:concept}(a), consists of two filters surrounding a non-reciprocal modulator/wavelength converter. This non-reciprocal converter is the focus of this work and the aspect which is demonstrated later in the paper. It utilizes two identical dual-active-cavity microring wavelength converters/modulators \cite{wade2015wavelength,gevorgyan2019efficient} and a microring phase shifter (an all-pass filter) \cite{chang2008tunable} to generate a sideband in the forward direction, shifted from the carrier by the RF frequency $\Omega$ due to constructive conversion/modulation from the two wavelength converters. In the backward direction, the two converters operate destructively and no light is converted to the sideband. The filters at either end of the non-reciprocal converter only pass the sideband, such that in the forward direction light can propagate through the isolator (frequency up-converted by $\Omega$) whereas in the backward direction no light can pass. The actions of each component in the non-reciprocal wavelength conversion section are shown in Fig.~\ref{fig:concept}(b) with a Mach-Zehnder interferometer analogue (each component is vertically aligned with its representative action in the inteferometer). In the forward direction, the non-reciprocal section converts between wavelengths ($\omega_1\to\omega_2$, $\omega_2\to\omega_1$) while in the backwards direction the incident light is unconverted ($\omega_1\to\omega_1$, $\omega_2\to\omega_2$).


To utilize this non-reciprocal wavelength converter as an isolator, we require only a single bandpass filter on either side, centered on either $\omega_1$ or $\omega_2$. If these bandpass filters are replaced by add-drop filters, the four resulting ports ($\omega_1$ and $\omega_2$ at both ends) can be used for a 4-port circulator. Perfect isolation is achievable even in the case of low modulation strength by exactly nulling out any conversion in the backward direction and choosing the input port to be at $\omega_2$ and the output port to be at $\omega_1$. In this work we demonstrate the non-reciprocal section and assume all light at the filtered frequencies is suppressed (up to 100 dB of filtering has been demonstrated in silicon photonic platforms \cite{gentry2018monolithic}) such that the ratio of wavelength conversion in the two directions is directly equal to the isolation.


A brief mathematical analysis of the system elucidates the non-reciprocal wavelength conversion and isolator performance. Consider the transmission matrix (T-matrix) representation of a dual-ring wavelength converter (derived using the relations in \cite{gevorgyan2020triply}) denoting the two resonance frequencies as the two ports \cite{wade2015wavelength}

\begin{equation}
    \overline{\overline{\mathbf{T}}}_\text{mod}=\begin{bmatrix}
    t_1 & -jt_2e^{-j\phi}\\
   -jt_2e^{j\phi} & t_1
    \end{bmatrix}
\end{equation}

\noindent where we have defined

\begin{equation}
    t_1=\frac{\delta\omega^2+16r_t^2-16r_er_t}{\delta\omega^2+16r_t^2}
\end{equation}
\begin{equation}
    t_2=\frac{4r_e\delta\omega}{\delta\omega^2+16r_t^2}.
\end{equation}

These relations assume input light at the lower supermode resonance ($\omega_1$), modulation at the frequency of the resonance splitting ($\Omega=2\mu$), and negligible contribution from oscillatory terms of the form $\Omega+2\mu$. Here, $r_0$ is the intrinsic single ring loss rate, $r_e$ is the bus coupling rate from the lower ring, and $r_t=r_0+r_e/2$ is the effective loss rate of a supermode. The term $\delta\omega$ denotes the amplitude of the modulation in terms of the resonance shift such that the instantaneous ring resonance (of a single ring) is $\omega_0(t)=\omega_0+(\delta\omega/2)\cos(\Omega t+\phi)$. An important point here is that the converted light at $\omega_2$ acquires the phase of the modulation $\phi$ whereas for light incident at $\omega_2$ the converted light at $\omega_1$ acquires the \emph{opposite} phase as a result of the inherent non-reciprocity of temporal modulation. 
The all-pass filter ($r_e\gg r_0$), functioning as a $\pi/2$ phase shifter of the sideband relative to the carrier, can be seen from evaluating the response of a ring resonator driven at $\pm\Omega$ and choosing the ring-bus coupling as $r_e=\Omega/(2+\sqrt{2})$. The T-matrix of the phase shifter in this case is

\begin{equation}
    \overline{\overline{\mathbf{T}}}_\text{ps}=e^{-j\pi/4}\begin{bmatrix}
    1 & 0\\
    0 & e^{j\pi/2}
    \end{bmatrix}.
\end{equation}

We can evaluate the performance in the forward and backward directions by sequentially multiplying these component T-matrices, where the first modulation has $\phi=0$ and the second modulator has $\phi=\pi/2$. The scattering matrix (S-matrix), comprising the T-matrices in both directions, is then

\begin{equation}
     \overline{\overline{\mathbf{S}}}_\text{nr}=e^{-j\pi/4}\begin{bmatrix}
    0 & 0 & t_1^2+t_2^2 & 0\\
    0 & 0 & 0 & j(t_1^2+t_2^2)\\
    t_1^2-t_2^2 & 2t_1t_2 & 0 & 0\\
    -2jt_1t_2 & j(t_1^2-t_2^2) & 0 & 0
    \end{bmatrix}.
\end{equation}

In accordance with the reciprocity relations (see e.g. \cite{jalas2013and}), an isolator requires a S-matrix  which is not symmetric (i.e. not invariant under the transpose operation, $\overline{\overline{\mathbf{S}}}_\text{nr}^T\neq  \overline{\overline{\mathbf{S}}}_\text{nr}$). The isolator configuration we use here converts light from $\omega_2$ at the input (represented by port 2 in the S-matrix) to $\omega_1$ at the output (represented by port 3 in the S-matrix). As can be seen, $ \overline{\overline{\mathbf{S}}}_{32}\neq  \overline{\overline{\mathbf{S}}}_{23}$, confirming the isolating properties of this system. The insertion loss can be identified as $| \overline{\overline{\mathbf{S}}}_{32}|^2$ and the isolation as $| \overline{\overline{\mathbf{S}}}_{32}|^2/| \overline{\overline{\mathbf{S}}}_{23}|^2$ (assuming perfect filtering at either end of the non-reciprocal section). Similarly, one could inject light at $\omega_1$ and convert to $\omega_2$ with identical performance.


\begin{figure}[t]
	\centering
	\includegraphics[width=.8\textwidth]{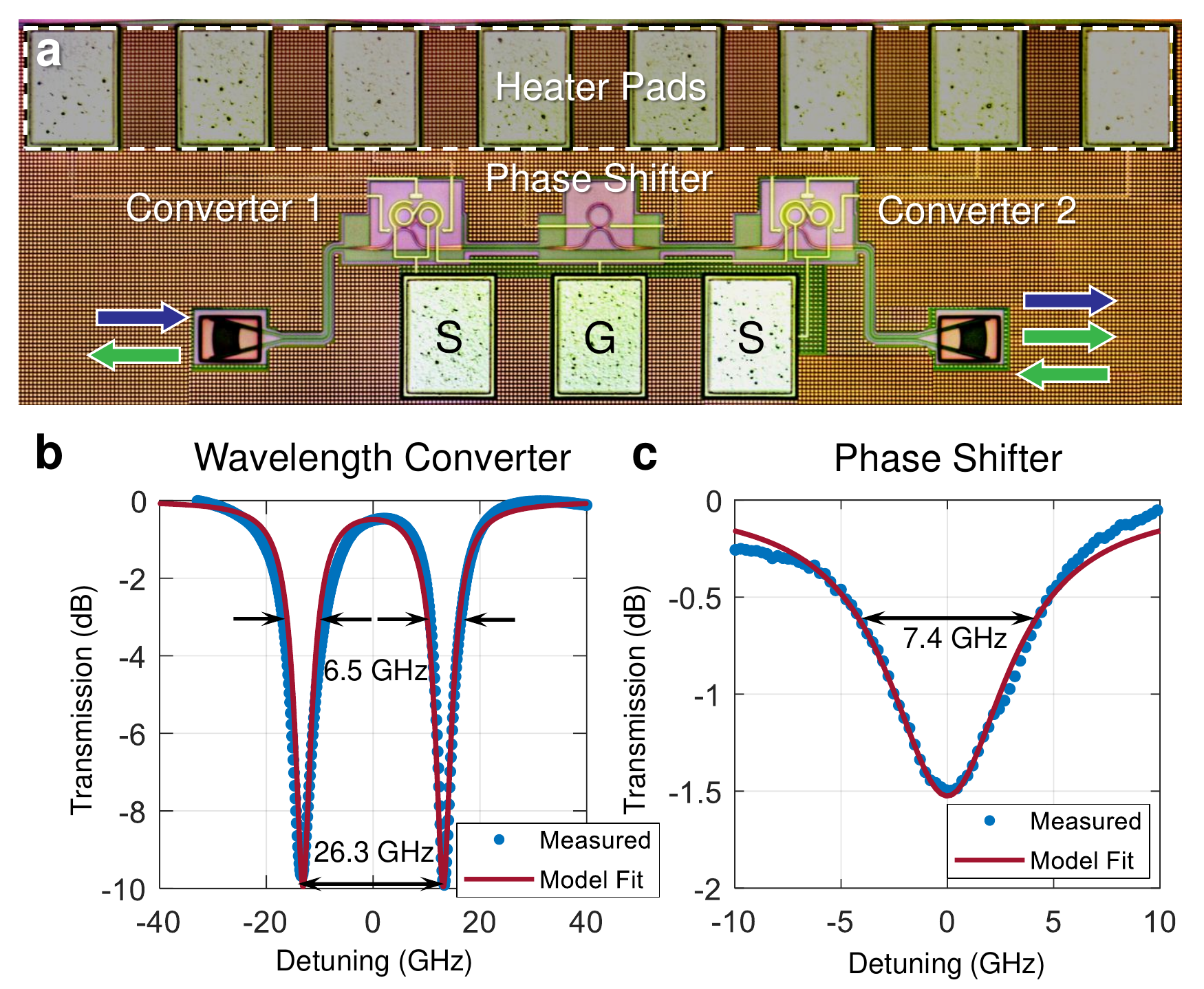}
	\caption{
	\textbf{Fabricated isolator and component performance.}
	\textbf{a} Image of fabricated isolator, thermal tuning pads (top), and RF pads (bottom).
	\textbf{b} Measured transmission of an individual wavelength converter without applied signal compared to a model fit (thermally detuned from phase shifter and second converter).
	\textbf{c} Measured transmission of the phase shifter ring compared to a model fit (thermally detuned from wavelength converters).}
	\label{fig:meas}
\end{figure}

We fabricated the non-reciprocal section of the isolator in an active silicon photonics process (IMEC). The fabricated device is shown in Fig.~\ref{fig:meas}(a), additionally showing the electrical pads used for electronic control and grating couplers used to optically address the device. We first characterized component performance to assess supermode splitting, intrinsic losses, and ideal alignment of the phase shifter ring to achieve the desired $\pi/2$ rotation. The measured transmission of both the wavelength converters and phase shifter ring are shown in Fig.~\ref{fig:meas}(b,c) respectively, where no modulation has been applied but thermal tuning is used to create a spectral separation between each component's transmission. The measurements indicate 26.3 GHz supermode splitting at which the electrical drive frequency should be set for maximum conversion efficiency. The linewidths of the resonant modes, here 6.5 GHz, determine the bandwidth of the converters, which in turn constrains the isolator bandwidth. The transmission of the phase shifter (all-pass filter) indicates a linewidth of 7.4 GHz.

\begin{figure}[t]
	\centering
	\includegraphics[width=.8\textwidth]{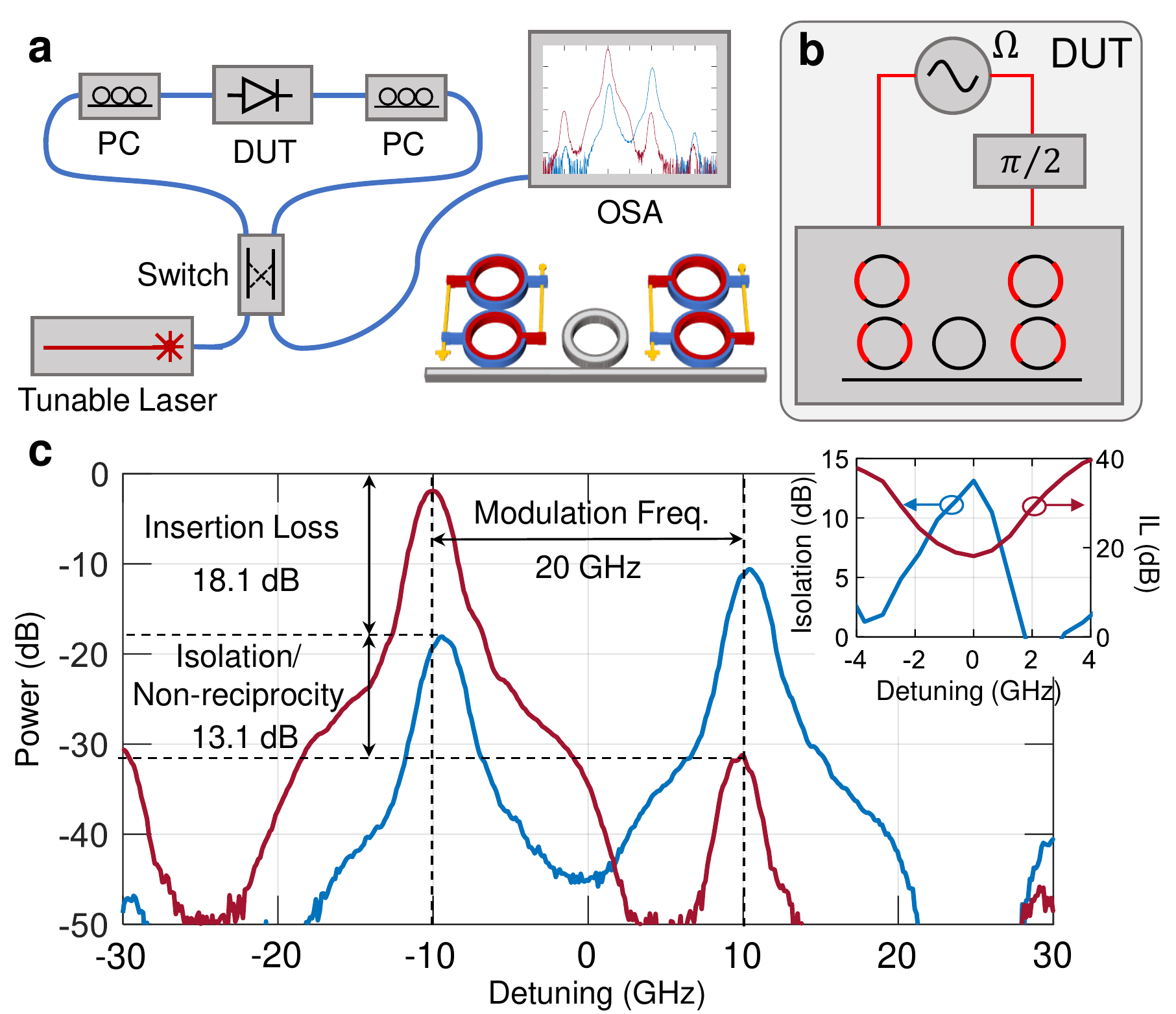}
	\caption{
	\textbf{Isolation demonstration.}
	\textbf{a} Experimental setup (based on that suggested in \cite{jalas2013and}); a switch enables testing of both forward and backward directions. OSA -- optical spectrum analyzer; DUT -- device under test (bottom right).
	\textbf{b} Schematic depiction of the DUT, where a single RF signal is applied to both converters (one phase-shifted by $\pi/2$). Not depicted: thermal tuning circuitry used to align ring resonances.
	\textbf{c} Demonstration of isolation, showing measured spectra in the forward (blue) and backward (red) directions. The carrier in the forward direction is aligned to the upper supermode resonance, and in the backward direction is aligned to the lower supermode resonance as discussed in the text. Inset: isolation (blue) and insertion loss (red) plotted vs. detuning from resonance.
	}
	\label{fig:iso}
\end{figure}

To demonstrate isolation, we use the setup shown in Fig.~\ref{fig:iso}(a) based on that suggested in \cite{jalas2013and}. A switch is used to allow measurement of transmission through the isolator in both directions without disconnecting any components. Excess insertion loss is measured by comparing to a reference value when the laser is off-resonance, which encompasses all losses external to the isolator. A tunable laser is used to input light at the upper (lower) supermode resonance in the forward (backward) direction, and an optical spectrum analyzer is used to compare carrier and sideband amplitudes. The electrical setup of the isolator is shown schematically in Fig.~\ref{fig:iso}(b), where we drive the isolator with a continuous-wave (CW) electrical signal at 20 GHz, limited by the available equipment. Improved conversion efficiency (insertion loss) can be expected when driving with a signal exactly matched to the mode splitting (26.3 GHz). A total power of -3 dBm is applied through signal-ground-signal (SGS) contacts, which after the power splitter and internal losses produces 0.21 V peak-to-peak voltage swing (-9.5 dBm) on each active cavity of the wavelength converters corresponding to $\delta\omega=1.3$ Grad/s based on the measured 1 GHz/V frequency shift. No DC bias is applied to the modulators. The $\pi/2$ phase shift of the RF signal to the second converter is achieved by introducing an extra length of RF cable to the second signal contact after a 50:50 splitter.

Using the thermal tuners, the corresponding mode resonances of the wavelength converters are aligned with each other and the microring phase shifter is tuned to produce the required $\pi/2$ phase difference.  Fig.~\ref{fig:iso}(c) depicts the measured spectra in the forward (blue) and backward (red) directions, demonstrating non-reciprocal mode conversion which results in isolation after filtering. These results are shown for light incident exactly at $\omega_2$ in the forward direction, and $\omega_2-\Omega$ in the backward direction. For this optimum case, the device shows a non-reciprocal response with 18.1 dB insertion loss and 13.1 dB isolation. We then measured the sideband power as a function of carrier detuning from the resonance ($\omega-\omega_2$ in the forward direction, $\omega-\omega_2-\Omega$ in the backward direction) to determine the variation of isolation and insertion loss with detuning. The measured 3 dB isolation bandwidth is approximately 2 GHz, within which the insertion loss remains below 20 dB.

\begin{figure}[t]
	\centering
	\includegraphics[width=.8\textwidth]{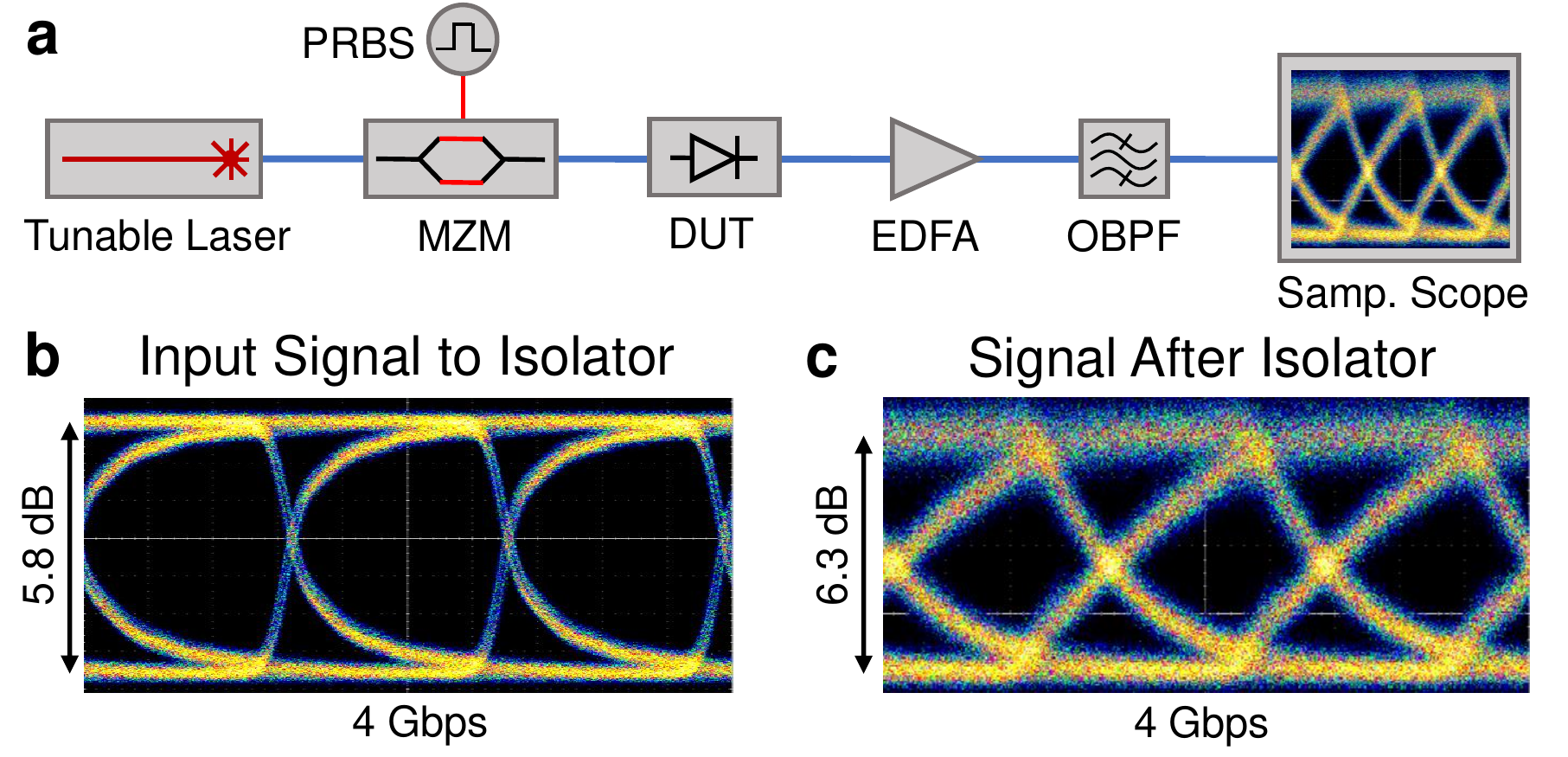}
	\caption{
	\textbf{Demonstration of data transmission through isolator.}
	\textbf{a} Experimental setup used for data transmission demonstration. PRBS -- pseudo-random bit sequence; MZM -- Mach-Zehnder modulator; DUT -- device under test; EDFA -- erbium-doped fiber amplifier; OBPF -- optical band pass filter.
	\textbf{b} Modulated data signal (after band pass filtering).
	\textbf{c} Data signal after transmission through the isolator.}
	\label{fig:data}
\end{figure}

This isolator is also suitable for use in wavelength division multiplexed communication links when the (identical) rings' free spectral range is matched to the channel spacing. For example, an isolator can be used to mitigate cross-talk between a transmitter and receiver which use the same wavelength and physical channels (e.g. optical fiber). To this end, we also demonstrate transmission of a 4 Gbps on-off keying (OOK) optical data signal through the isolator (Fig.~\ref{fig:data}). The experimental setup used is shown in Fig~\ref{fig:data}(a) where the isolator configuration is identical to that used for demonstration of isolation (Fig.~\ref{fig:iso}). A bit sequence is encoded on a CW tone using a fiber-coupled modulator driven by a bit pattern generator (Picosecond Pulse Labs 12050). The modulated signal is shown in Fig.~\ref{fig:data}(b) after band pass filtering. We then pass the modulated signal (without filtering) through the isolator and use a fiber amplifier to recover signal power. The isolated signal is shown in Fig.~\ref{fig:data}(c) after band pass filtering, demonstrating the isolator's ability to be used for communication applications.


The demonstrated isolation of 13 dB presents, to the best of our knowledge, a new record in silicon photonics amongst modulation-based approaches. That it is still far from the 20-40 dB needed in a good isolator is an indication of the difficulty in implementing this function in pure silicon. The present design can reach such a performance with only optimizations of the design. The main advantage of our approach is the use of resonance to replace the multi-mm long delay line required in other approaches to a single, {\textmu}m-scale microring resonator enabled by the use of single sideband modulation. The compact layout is comparatively immune to fabrication variations such as waveguide thickness variations that result in imperfect cancellation of the two modulators (low isolation) in previous designs \cite{doerr2014silicon,lira2012electrically}, requiring only standard thermal tuning to ensure the rings are properly aligned. This increased isolation comes at the cost of increased insertion loss due to low conversion efficiency, as well as restricted operation bandwidth arising from resonance. Notably, many applications can function within this bandwidth restriction, particularly laser isolation. Even tunable lasers operating with more than 100 nm bandwidth can utilize resonant isolators so long as the resonances can be tuned to track the instantaneous wavelength, like the isolator demonstrated here. One remaining limitation of such an approach is the response time of the tuning; with thermal tuning such as that demonstrated here, {\textmu}s-scale response times can be expected \cite{sun201645}.

Several potential improvements could drastically decrease insertion loss and increase isolation. The undesirably high insertion loss demonstrated here is a result of inefficient sideband generation. Various methods of increasing conversion efficiency in microring resonators can be used, such as an RF resonance at the modulation frequency \cite{gevorgyan2020triply}, increasing the ring quality factor (at the cost of optical bandwidth), or even hybrid integration of materials with high electro-optic coefficients \cite{kieninger2018ultra}. For example, the voltage applied here combined with the demonstrated V$\pi$ in \cite{kieninger2018ultra} would achieve approximately maximum conversion corresponding to up to 78\% conversion efficiency (1 dB insertion loss). In addition to undesirably high insertion loss, the current demonstration has an isolation limited by imperfect cancellation of the two sidebands in the backward direction, which can result from a phase shift differing from $\pi/2$ or imbalance of sideband amplitudes. More precise control of the modulation strength, modulation phase, and phase shifter resonance alignment would enable increased isolation. On-chip feedback control, such as that demonstrated in \cite{sun201645} for thermal control of ring resonances, could lock onto the light transmitted in the backward direction at the rejected port of the filter and continually optimize isolation.

In conclusion, this paper presents a new isolator concept for silicon photonics which is compact, linear, silicon only, and free of pump lasers. We demonstrate the isolator in an active silicon photonic process and measure up to 13 dB of isolation (3 dB bandwidth of 2 GHz) with 18 dB insertion loss in a first demonstration. Additionally, we show transmission through the isolator of a 4 Gbps data signal while maintaining an open eye. We expect that this isolator demonstration will renew interest in silicon photonic isolators operating without the need for magneto-optic materials towards the goal of black box isolators being an everyday component in advanced photonic foundries.

\noindent \textbf{Funding Information} National Science Foundation Graduate Research Fellowship Grant (1144083); Packard Fellowship for Science and Engineering (2012-38222).


\bibliographystyle{ieeetr}
\bibliography{biblio}

\end{document}